\documentclass[useAMS,usenatbib]{mn2e}
\usepackage{psfig, epsf, epsfig}

\title[A ULX microquasar in NGC\,5408?]{A ULX microquasar in NGC\,5408?}
\author[R. Soria et al.]{R. Soria$^{1}$\thanks{E-mail:
rsoria@cfa.harvard.edu (RS)}, 
R. P. Fender$^{2}$, D. C. Hannikainen$^{3}$, A. M. Read$^{4}$ and 
I. R. Stevens$^{5}$\\
$^{1}$Harvard-Smithsonian Center for Astrophysics, 
	60 Garden st, Cambridge, MA 02138, USA\\
$^{2}$School of Physics \& Astronomy,
University of Southampton, SO17 1BJ, UK\\
$^{3}$Observatory PO Box 14, 00014 University of Helsinki,
Finland \\
$^{4}$Department of Physics \& Astronomy,
University of Leicester, LE1 7RH, UK\\
$^{5}$School of Physics and Astronomy, University of Birmingham,
B15 2TT, UK}
\begin{document}

\date{Received 2006 Jan 09; accepted 2006 Feb 28}

\pagerange{\pageref{firstpage}--\pageref{lastpage}} \pubyear{2005}

\maketitle

\label{firstpage}

\begin{abstract}
We studied the radio source associated with the ultraluminous 
X-ray source in NGC\,5408 ($L_{\rm X} \approx 10^{40}$ erg s$^{-1}$). 
The radio spectrum is steep (index $\approx -1$), 
consistent with optically-thin synchrotron emission, 
not with flat-spectrum core emission. 
Its flux density ($\approx 0.28$ mJy at 4.8 GHz, 
at a distance of 4.8 Mpc) was the same in the March 2000 
and December 2004 observations, suggesting 
steady emission rather than a transient outburst.
However, it is orders of magnitude higher than expected 
from steady jets in stellar-mass microquasar.
Based on its radio flux and spectral index, 
we suggest that the radio source is either an unusually bright 
supernova remnant, or, more likely, a radio lobe powered 
by a jet from the black hole.
Moreover, there is speculative evidence that the source is marginally 
resolved with a radius $\sim 30$ pc. A faint H\,{\footnotesize II} 
region of similar size appears to coincide with the radio 
and X-ray sources, but its ionization mechanism remains unclear. 
Using a self-similar solution for the expansion 
of a jet-powered electron-positron 
plasma bubble, in the minimum-energy approximation,  
we show that the observed flux and (speculative) size are consistent 
with an average jet power $\approx 7 \times 10^{38}$ erg s$^{-1}$ 
$\sim 0.1 L_{\rm X} \sim 0.1 L_{\rm Edd}$, 
an age $\approx 10^5$ yr, a current velocity 
of expansion $\approx 80$ km s$^{-1}$. We briefly discuss 
the importance of this source as a key to understand the balance 
between luminosity and jet power in accreting black holes.
\end{abstract}


\section{Introduction}

Ultra-luminous X-ray sources (ULXs) are point-like, 
accreting X-ray sources with apparent isotropic 
luminosities $\ga$ a few $10^{39}$ erg s$^{-1}$, that is, 
greater than the Eddington limit of a stellar-mass 
black hole (BH). Their nature remains unexplained. 
In particular, for the large majority of ULXs, 
two fundamental questions remain unsolved:
whether the emission is isotropic or 
beamed towards the observer (either as 
a mild geometrical beaming, or as 
a collimated, relativistic jet); 
and whether the accreting compact objects are young, 
perhaps formed in recent starburst episodes, or are 
instead old relics of a primordial 
stellar population.

Studying the ULX counterparts at other wavelengths 
may offer clues on their ages, mechanisms of formation, 
isotropic luminosities (through their 
effect on the interstellar medium) and sources of fuel.
In particular, it has recently become clear 
that there is a strong connection between 
X-ray and radio properties of accreting BHs.
This connection has been successfully applied 
to BH X-ray binaries in our Galaxy, providing 
a physical interpretation to their state transitions 
(Fender, Belloni \& Gallo 2005).
Similar studies for accreting BHs in nearby galaxies 
(e.g., Fender, Southwell \& Tzioumis 1998)
are, however, hampered by the faintness 
of the radio counterparts.

The ULX in the nearby ($d = 4.8$ Mpc: 
Karachentsev et al.~2002) dwarf starburst 
galaxy NGC\,5408 (henceforth, X-1) 
is an ideal test case for our study.
Its position is known to $\la 0\farcs5$, 
from a series of {\it Chandra} observations.
The source is among the brightest 
ULXs in the local Universe, with 
an isotropic X-ray luminosity $\approx 1 \times 10^{40}$ 
erg s$^{-1}$ in the $0.3$--$12$ keV band.
Always detected in an active state also 
by {\it Einstein}, {\it ROSAT} and {\it 
XMM-Newton}, X-1 has shown a pattern of 
flux variability over a few weeks or months 
(Kaaret et al.~2003), 
as well as on shorter timescales (Soria et al.~2004), 
inconsistent with an X-ray supernova remnant (SNR).
Thus, there is no doubt that the X-ray emission 
is due to accretion onto a compact object.
Most importantly, it is one of the very few 
ULXs with a radio counterpart (Kaaret et al.~2003).
The possibility that it is a background object 
(radio galaxy or quasar) can be ruled out 
due to the extremely high X-ray to optical 
flux ratio (Kaaret et al.~2003); a background blazar 
is also very unlikely, because their radio spectral indices  
are much flatter than determined for this source 
(Kaaret et al.~2003 and Section 2 of this work).

Two alternative scenarios have been suggested 
for X-1 (Kaaret et al.~2003; Soria et al.~2004). 
It could be a stellar-mass microblazar (i.e., 
a microquasars with its relativistic jet oriented 
along our line of sight). In this case, 
its extreme brightness would be due to Doppler boosting; 
its radio emission would be synchrotron from the jet; 
its X-ray spectrum could be produced by synchrotron 
self-Compton or external Compton scattering 
of low-energy photons (including UV/optical photons 
from the disk or the donor star) by the relativistic 
jet (K\"{o}rding, Falcke \& Markoff 2002; 
Georganopoulos, Aharonian \& Kirk 2002; 
Georganopoulos \& Kazanas 2003). 
Alternatively, the X-ray emission could be isotropic, 
implying a mass $\ga 100 M_{\odot}$ for the accreting BH, 
if the Eddington limit is not persistently 
violated. Applying conventional phenomenological models, 
the $0.3$--$10$ keV spectrum is dominated by 
a rapidly-varying power-law component 
with photon index $\Gamma \approx 2.7$, 
plus a non-varying soft thermal component 
or ``soft excess'' ($kT \approx 0.13$ keV), 
often found in ULXs.

We re-observed the system in the radio, 
with a more extended spectral coverage 
(four instead of two bands),
to measure its spectral index, and  
to determine whether there has been 
a long-term variation in the radio flux 
compared with the previous observations 
in March 2000 (Stevens, Forbes \& Norris 2002).
Using these new results, 
we want to determine whether the radio 
emission is coming directly from  
the accreting BH (for example, from a relativistic jet) 
or is instead only spatially associated with it 
(for example, it may come from an underlying SNR). 
In either case, we shall obtain important information
on the geometry of emission or on the immediate 
environment of the ULX.
More generally, we want to determine whether 
the associated radio and X-ray emissions in X-1 
are consistent with the correlations found in 
Galactic BH binaries, 
as a function of spectral state (Fender et al.~2004).

\section{ATCA radio study}

We observed NGC\,5408 on 2004 December 9--11, 
with the Australia Telescope Compact Array (ATCA), 
located at Narrabri, New South Wales. The array 
was in the 6-km configuration. 
Two sets of observations were carried out: 
simultaneously at 1.344 GHz (20 cm) and 2.638 GHz (13 cm); 
and simultaneously at 4.840 GHz (6 cm) and 6.208 GHz 
(5 cm)\footnote{We chose to observe at 5 cm, rather 
than the more standard 3 cm, because of 
severe weather conditions, which made it impossible 
to preserve phase coherence at higher frequencies.}, 
with a continuum bandwidth of 128 MHz.
Observations of the target galaxy were alternated 
with those of the primary flux calibrator 
PKS\,B1934$-$638, and of the secondary phase calibrator 
PKS\,B1320$-$446. In total, we had about 13 hr spent 
on source for each of the two configurations.

The data were processed and analysed using the {\footnotesize MIRIAD}
package (Sault, Teuben \& Wright 1995). Images were cleaned with 1000
iterations of gain 0.1. The clearest image of the radio counterpart to
the ULX is available at 4.8 GHz (Figure 1), in which the source is clearly
resolved from the nearby starburst and has a flux
density of $0.28 \pm 0.07$ mJy (consistent with Kaaret et
al. 2003). The best-fit radio coordinates of the source associated
with the ULX, obtained from the peak of the emission in the 4.8 GHz
map, is R.A. = $14^h03^m19^s.65$, 
Dec = $-41^{\circ}22\arcmin 58\farcs7$ with an error 
$\approx 0\farcs3$.  
This is consistent (within $0\farcs2$) with the X-ray
position of the ULX (R.A. = $14^h03^m19^s.63$, 
Dec = $-41^{\circ}22\arcmin 58\farcs7$; 
Kaaret et al.~2003), 
which we also re-checked by analysing the {\it Chandra}
observations available in the public archive.

From the images at other wavelengths it was immediately apparent that
the relative faintness of the target, combined with its proximity to
the starburst region, was resulting in strong uncertainties in the
measured flux density. In order to quantify this, we have varied the
beam size at each frequency by adjusting Briggs' ROBUST parameter
between -2 and +2, corresponding to a smooth transition from uniform
(highest angular resolution, lowest signal-to-noise) to natural
(lowest angular resolution, highest signal to noise). For each
resultant map we have fit a point source model, with the coordinates
fixed to those established from the 4.8 GHz map. The variation of the
apparent source flux as a function of beam size, for each frequency, is
presented in Figure 2. Beam size is clearly a strong effect
at 1.3 and 2.4 GHz, less so at 4.8 GHz and has no significant effect
at 6.2 GHz: hence, it is hard to confidently ascribe
a well measured flux density to the radio counterpart 
of the ULX, particularly at low frequencies. 
A 4.8-GHz radio flux density of $(0.28 \pm 0.07)$ mJy 
corresponds to a radio luminosity of 
$(3.7\pm 0.9) \times 10^{34}$ erg s$^{-1}$. 
As noted above, the measured
flux is consistent with the 4.8-GHz flux inferred from the March 2000
ATCA observations (Kaaret et al.~2003; Stevens et al.~2002),
suggesting that the source has not varied significantly between the 
two epochs.




\begin{figure}
\epsfig{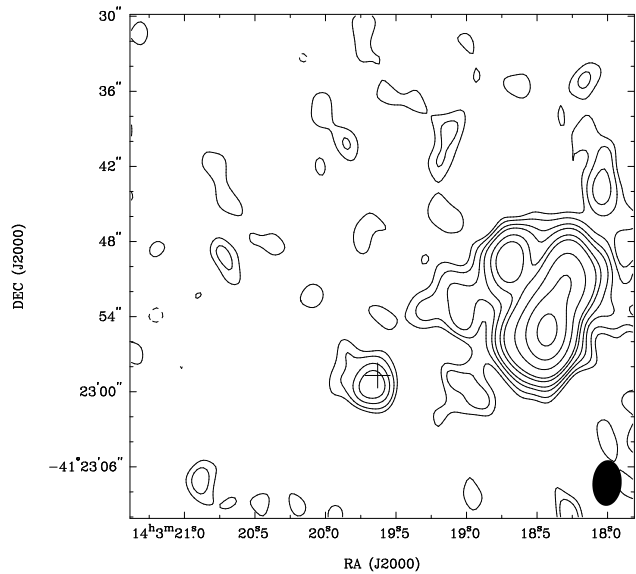}
\caption{ATCA radio map at 4.8 GHz (6 cm). 
The X-ray position of the ULX is marked with a cross.
The integrated 4.8-GHz flux from the ULX counterpart 
is $\approx 0.28$ mJy.}
\end{figure}


\begin{figure}
\epsfig{figure=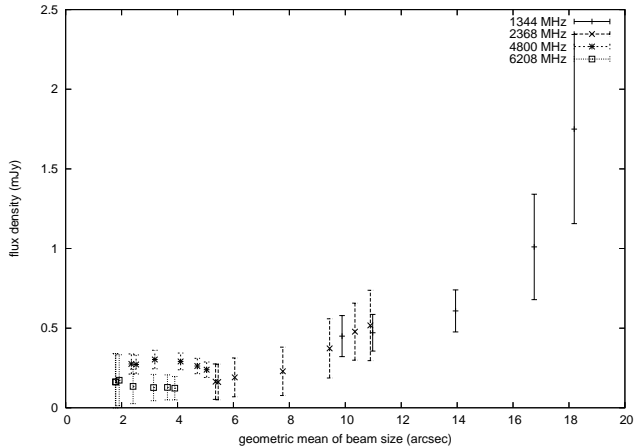, width=6cm, angle=270}
\caption{Changes in the apparent source flux as a function 
of beam size, in the four radio bands. At low frequencies, 
the source is contaminated by extended emission 
from nearby star-forming regions. See Section 2 for details.}
\end{figure}

\begin{figure}
\epsfig{figure=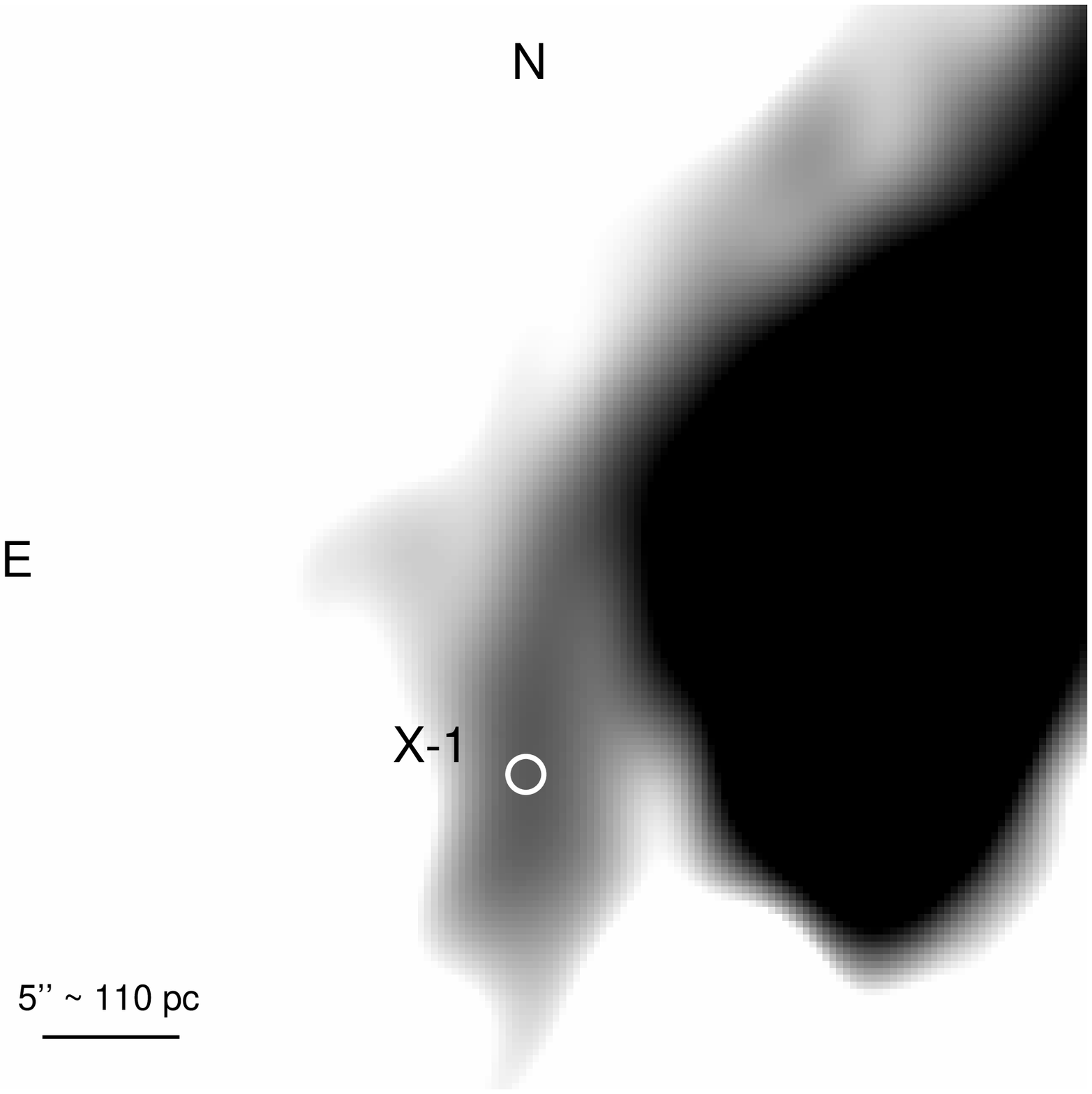, width=8.6cm}
\epsfig{figure=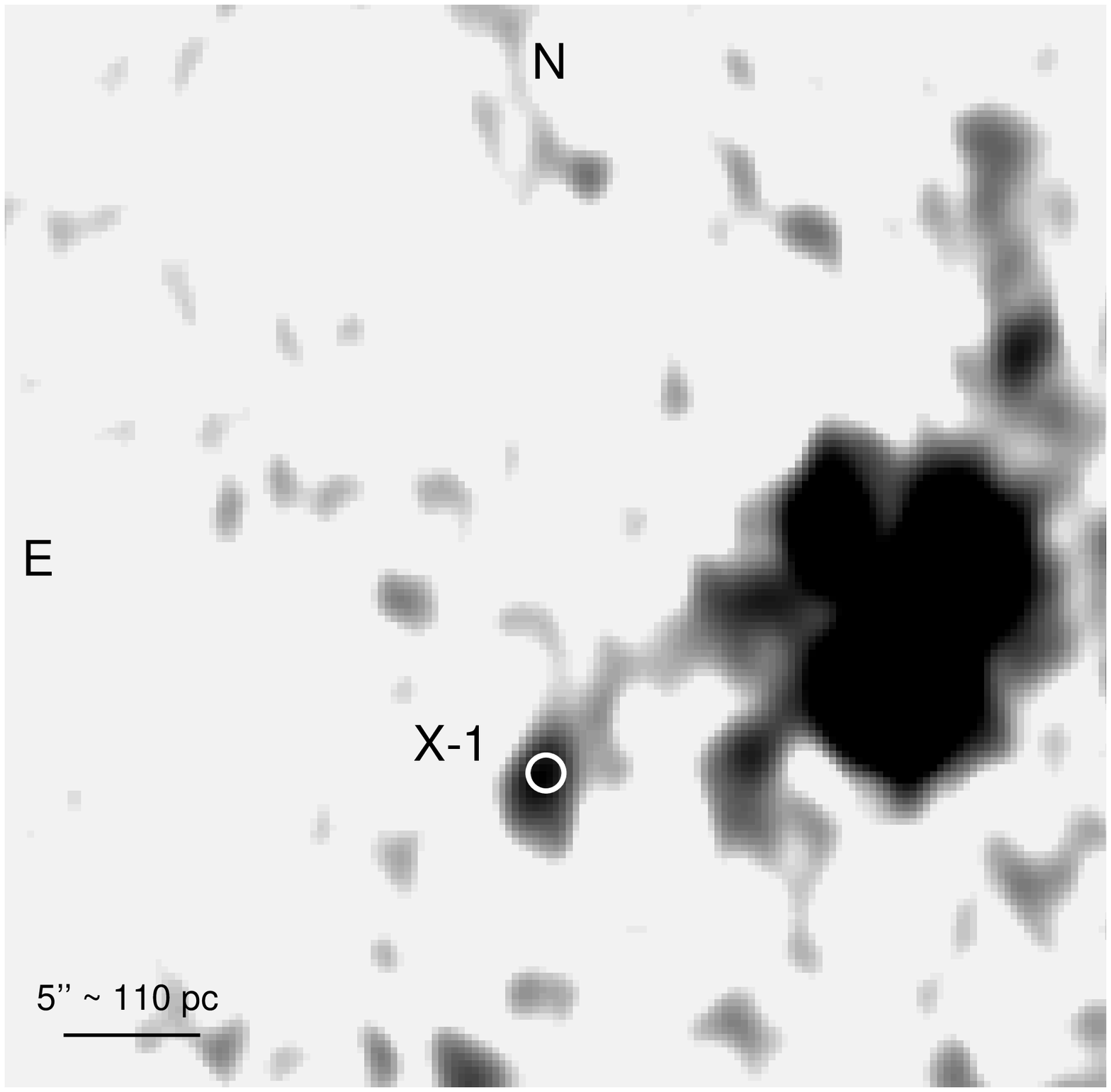, width=8.6cm}
\caption{Top panel: radio map of the field 
obtained by combining the 1.3 and 2.4 GHz data.  
The {\it Chandra} position of the ULX is marked 
with an error circle of radius $0\farcs6$. 
Bottom panel: radio image obtained by combining 
the 4.8 and 6.2 GHz maps.}
\end{figure}

\begin{figure}
\epsfig{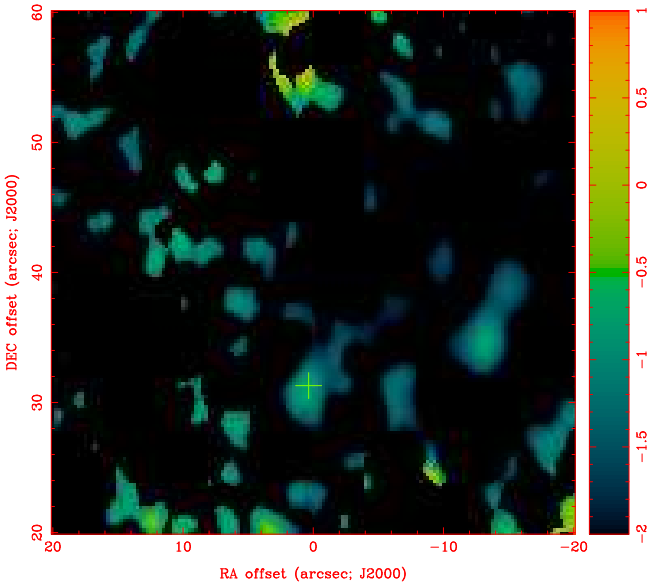}
\caption{Ratio of the high-frequency and low-frequency maps, 
showing that the ULX counterpart has a steep spectral index 
($\alpha_{\rm R} \approx -1$). The scale and orientation 
of the image is the same as in Figure 3; the ULX position is marked 
with a cross.}
\end{figure}

Between the highest two frequencies, the marginal detection at 6.2 GHz
(in the naturally weighted maps, the point source fit at the position
fixed from the 4.8 GHz image is $0.12 \pm 0.07$ mJy) suggests at face
value a very steep spectrum with spectral index 
$\alpha_{\rm R} \leq -3$\footnote{Throughout the paper, the spectral 
index is defined such that $F_{\nu} \sim \nu^{\alpha}$, 
and we use $\alpha_{\rm R}$ or $\alpha_{\rm X}$ 
to avoid confusion between the radio and X-ray spectral slopes.
Using the notation more commonly adopted in X-ray studies, 
the X-ray photon index $\Gamma \equiv -\alpha_{\rm R} + 1$.}. 
This is an extreme value for synchrotron emission: 
it may suggest that the break due to ageing of the electron 
populations is located between those two bands. 
We shall discuss---and essentially rule out---this 
possibility in Section 3.6. Alternatively, it may be taken 
as a hint that we are starting to resolve the source 
at those higher frequencies, which would imply 
a characteristic size $\approx 3\arcsec \times 2\arcsec$, 
or $\approx 50 \times 70$ pc at the assumed distance 
of NGC\,5408.
At lower frequencies, contamination by the flux from
the starburst starts to dominate. This makes determining the radio
spectrum associated with the ULX problematic.

In another approach to estimate the radio spectral index at the ULX
position and in the surrounding region, we obtained  
radio images of the field by combining the
1.3 and 2.4 GHz maps (Figure 3, top panel) 
and the 4.8 and 6.2 GHz maps (Figure 3, bottom
panel).  The radio source associated with the ULX clearly 
stands out as a discrete (and possibly resolved) blob 
in the combined high-frequency map, with hints of
emission trailing back to the nearby bright starburst region. In
the low-frequency map, there is clearly radio emission at the ULX
location, apparently as the end of an elongated region emanating from
the starburst complex, but our spatial resolution is not sufficient
for a more detailed analysis. We then took the ratio of the
low-frequency and high-frequency maps (Figure 4).  From this, we note
that the bright starburst region is clearly strongly resolved out on
its outer edges, which is why the spectral index in these regions
appears steeper than $-2$. Instead, both the core of the starburst
region and the site of the ULX are not resolved out, and have spectral
indices $\approx -1$ between these two combined bands. 
So, while a qualitative approach
based on the measured flux densities at four frequencies for the ULX
is problematic, there is strong evidence that the
radio emission at the site of the ULX has a steep spectral index, 
consistent with optically-thin synchrotron
emission, and in agreement with the March 2000 
observations (Kaaret et al.~2003). 
Incidentally, the flux-ratio map also suggests that the radio emission in
the core of the starburst region is dominated by SNRs (steep spectrum)
rather than free-free emission from H\,{\footnotesize{II}} regions
(flat spectrum).

\section{Discussion}

\subsection{Steady jet in a low/hard state?}

Given that the evidence for a spatially resolved 
radio source is still only marginal, we shall 
consider both the scenario of a point-like 
source (core emission) and that of an extended source.
We start by considering the possibility 
that the radio emission is directly produced 
by the accreting BH, on spatial 
scales comparable to the accretion disk. Radio emission 
from Galactic BH X-ray binaries is known to occur---with 
different physical properties---in two 
spectral states: the ``low/hard'' state (LHS) 
and the ``very-high'' state (VHS) (Fender et al.~2004).

The LHS is characterised by a steady jet, 
with moderately low bulk Lorentz factor ($\gamma \la 1.4$: 
Fender et al.~2004; Gallo, Fender \& Pooley 2003).
In a steady state, there is a ``fundamental-plane'' 
correlation between radio luminosity, X-ray luminosity 
and BH mass (Falcke, K\"{o}rding \& Markoff 2004; 
see also Merloni, Heinz \& DiMatteo 2003; Fender et al.~2004), 
of the form 
\begin{equation}
L_{\rm X} \propto L_{\rm R}^m 
M_{\rm BH}^{\alpha_{\rm X} -m\alpha_{\rm R}},
\end{equation} 
where 
\begin{equation}
m = \frac{(17/12) - (2/3)\alpha_{\rm X}}{(17/12)-(2/3)\alpha_{\rm R}}
\end{equation}
(Markoff et al.~2003).
Taking this relation at face value---or, 
equivalently, using the empirical relation 
$\log L_{\rm R} = 0.60 \log L_{\rm X} + 0.78 \log M + 7.33$ 
(Merloni et al.~2003)---we would infer a BH mass 
$\sim 10^4 M_{\odot}$, well in excess of any estimates of 
other ULX masses, including the strong 
intermediate-mass BH candidate  
in M\,82 (Strohmayer \& Mushotzky 2003; 
Fiorito \& Titarchuk 2004).

However, there are at least two reasons why we 
rule out this result as unphysical.
Firstly, the observed radio spectral index 
($\alpha_{\rm R} \approx -1$) is too steep 
to be consistent with an LHS steady jet, which 
is typically optically thick (core-dominated), 
with flat or slightly 
inverted radio spectrum (Fender et al.~2004).
Secondly, the X-ray spectrum (Soria et al.~2004) 
is itself inconsistent with an LHS. The X-ray spectrum 
in a LHS is dominated by a power-law 
of spectral index $\alpha_{\rm X} \approx -0.5$ 
(i.e., photon index $\Gamma \approx 1.5$ 
as more commonly defined in X-ray literature).
Instead, the observed X-ray spectrum of X-1 is much steeper 
(softer), with $\alpha_{\rm X} = -1.7\pm 0.2$ 
(i.e., $\Gamma \approx 2.7$).
We conclude that the radio emission is not consistent 
with a steady jet from an accreting BH in the LHS.


\subsection{Flaring jet in the very high state?}

The X-ray spectrum (Soria et al.~2004) 
is consistent with the classical  
VHS of Galactic X-ray binaries, also called 
the ``steep power-law'' state, 
using the revised state classification of 
McClintock \& Remillard (2003). 
Steady jets with a flat or inverted radio spectrum 
are sometimes observed in the ``hard'' 
sub-division of the VHS: this can be considered 
a high-luminosity extension of the LHS 
discussed (and ruled out) in Section 3.1.
Stronger radio flares are observed in Galactic sources 
at the transition between the ``hard'' 
and ``soft'' subdivisions, as the 
sources become softer and cross the ``jet line'' 
(see Figure 7 of Fender et al.~2004).
These flares are interpreted as discrete relativistic 
radio ejections, with a higher bulk Lorentz factor 
than in an LHS jet ($\gamma \sim$ a few: Gallo 
et al.~2003). One explanation for the steep-spectrum 
radio emission is optically-thin synchrotron, 
from internal shocks in the jet, consequence 
of the sudden increase in the Lorentz factor 
during the state transition (Fender et al.~2004).

Well-known examples of Galactic BH binaries
with a radio flaring behaviour in their 
VHS include GRS 1915$+$105, GRO J1655$-$40, and 
XTE J1550$-$564 (Table 1). If moved to the 
distance of NGC\,5408, even at their peaks 
they would appear two to three orders 
of magnitude fainter than the radio counterpart of X-1.
Moreover, the radio to X-ray flux ratio 
is at least an order of magnitude higher 
in X-1, compared to those Galactic sources (Figure 5).
Again, one might suggest that the enhanced 
radio luminosity, and radio-to-X-ray flux ratio 
are due to a much higher BH mass in X-1.

\begin{table*}
\begin{tabular}{lcccccc}
\hline
Source  & $M_{\rm BH}$ & Date & Peak 5-GHz flux 
	& $L_{\rm J}/L_{\rm Edd}$ & $L_{\rm X}/L_{\rm Edd}$ & Ref.\\
 & ($M_{\odot}$) & (MJD) & (mJy at 4.8 Mpc) & & &\\
\hline
GRS 1915$+$105 (flr) & 14 & 50750 & $1.7 \times 10^{-3}$ & 0.6 & 1.1 & F99\\
GRS 1915$+$105 (osc) & 14 & 50750 & $2.6 \times 10^{-4}$ & 0.05 & 1.1 & F99\\
XTE J1748$-$288 & 7 & 50980 & $1.7 \times 10^{-3}$ & 1.9 & 0.1 & B06\\
V4641 Sgr    & 9 &   51437 & $1.2 \times 10^{-3}$ & 0.8 & 4 & H00, O01\\
GRO J1655$-$40 & 7 & 49580 & $1.1 \times 10^{-3}$ & 1.0 & 0.1 & HR95\\
XTE J1550$-$564 & 9 & 51077 & $2.0 \times 10^{-4}$ & 0.3 & 0.5 & W02\\
XTE J1859$+$226 & 7 & 51467 & $7.8 \times 10^{-5}$ & 0.2 & 0.2 & B02\\
\hline
\end{tabular}
\caption{Parameters for the brightest radio outbursts 
in Galactic BH candidates. All jet events listed here 
occurred in the very-high-state, 
with both the jet power and the X-ray luminosity estimated 
to be $\sim$ Eddington. Data from Fender et al.~2004; 
other references: F99 = Fender et al.~1999; 
B02 = Brocksopp et al.~2002; B06 = Brocksopp et al., in prep.; 
H00 = Hjellming et al.~2000; O01 = Orosz et al.~2001; 
HR95 = Hjellming \& Rupen 1995; W02 = Wu et al.~2002.}
\end{table*}

However, even if X-1 is interpreted 
(based on its X-ray properties) as a BH binary in 
the VHS, quantitative comparisons with radio-flaring 
Galactic BHs are highly problematic.
The VHS is a short-lived transient state, 
and a major ejection is a brief (rise time 
$\la 1$ day), sporadic event (once every 
few months or years), as 
the source undergoes a hard to soft transition, 
associated with a change in the accretion disk 
inner radius.
In our case, the radio flux was the same 
on the two occasions in which the source 
was observed, almost five years apart.
Catching X-1 during a major flare once would be 
a remarkable coincidence; observing it 
twice in exactly the same state would 
be too unlikely.
Moreover, there is no evidence of any X-ray 
spectral state transitions over a series of {\it Chandra} 
and {\it XMM-Newton} observations spanning 
the years 2001--2004. Hence, it is more 
likely that the observed radio emission is 
a persistent feature of this source.

The only marginally viable alternative is that 
the source is persistently in the VHS, 
oscillating in a ``dip-flare'' cycle like 
GRS 1915$+$105, crossing the jet line back and forth 
in the process (Fender et al.~1999; 
Fender \& Belloni 2004 and references therein); 
this state is characterized by moderate X-ray and radio 
flaring on timescales of a few hundred seconds.
Repeated flaring on similar timescales 
is indeed observed in the power-law component 
of the X-ray spectrum of X-1 (Soria et al.~2004).
Finding the source in that same dip-flare cycle 
in 2000 and 2004 (and in a similar X-ray spectral 
state consistent with the VHS in all observations 
between those two epochs) could perhaps suggest 
that this ULX has a different spectral-state
behaviour than classical Galactic X-ray binaries.

The peak radio flux density of GRS 1915$+$105 in 
the oscillatory state is three orders of magnitude 
lower than the radio flux of X-1, scaled to the same 
distance (Table 1). Taken at face value, 
the intensity of the radio emission from X-1 
in 2000 and 2004 make this ULX very peculiar, 
in comparison with typical Galactic BH X-ray binary, 
in which the radio emission is believed 
to come from a compact jet (Figure 5).
However, we must also consider the possibility that 
the radio source associated to X-1 
is enhanced by Doppler boosting, as we shall discuss 
in Section 3.3.

\begin{figure}
\epsfig{figure=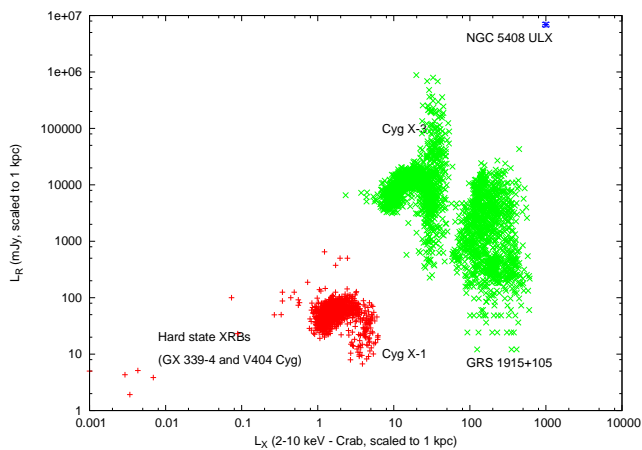, angle=270, width=8.5cm}
\caption{Comparison between the X-ray and radio fluxes 
from NGC\,5408 X-1 and from some of the brightest 
Galactic microquasars, either in the LHS (steady jet) 
or in the VHS (flarings). In the on-line version of this plot, 
LHS datapoints are plotted in red, VHS in green.
NGC\,5408 X-1 is a few orders 
of magnitude brighter, especially in the radio band. 
However, we argue that the comparison is not relevant, because 
the radio emission from the X-1 location comes either 
from an underlying SNR, or, much more likely, from a radio lobe.}
\end{figure}

\subsection{Relativistically-beamed emission?}

Relativistic Doppler boosting is a possible mechanism to reconcile 
the brightness of X-1 with those of Galactic BHs, 
as originally suggested by Kaaret et al.~(2003).
The amplification of the rest-frame emission due 
to Doppler boosting $\sim \delta^{k-\alpha_{\rm R}}$,
where $\delta = \gamma^{-1}(1-\beta \cos \theta)^{-1}$ 
and $2 \leq k \leq 3$ depending on whether it is a continuous 
jet or sporadic ejections.
Having already ruled out (Section 3.2) both a steady jet in the LHS 
and a single major flare, we only need to estimate 
the Doppler boosting required for a persistent dip-flare cycle.
In GRS 1915$+$105, the Lorentz factor $\gamma$ in this state 
is $\ga 2$ and perhaps as high as $5$ (Fender et al.~2004); 
at a viewing angle $\approx 65^{\circ}$, the observed $5$-GHz flux 
is $\approx 50$ mJy at 11 kpc, corresponding to 
$\approx 0.3$ $\mu$Jy at 4.8 Mpc. To enhance this value 
by a factor $\approx 10^3$, a viewing angle $\theta \la 20^{\circ}$ 
(for $k = 3$) or $\theta \la 15^{\circ}$ 
(for $k = 2$) is required, for a source identical 
to GRS 1915$+$105 and $\gamma = 5$. 
Allowing for an intrinsic radio luminosity 
an order of magnitude stronger than in GRS 1915$+$105 
(justified by the fact that the apparent X-ray luminosity 
is also a factor of 10 higher), only $\theta \la 30^{\circ}$ 
(for $k = 3$) or $\theta \la 25^{\circ}$ (for $k = 2$) are required.
As an aside, we note that in order to allow 
for persistent dip-flare ejections with Lorentz factors 
$\approx 5$, the microquasar would have 
to reside in a large low-density bubble, otherwise 
the jet-ISM interaction would rapidly reduce 
the jet speed.

For the X-ray emission, strong beaming is disfavoured 
by the presence of a soft, non-variable thermal component 
at $kT_{\rm bb} = 0.13 \pm 0.01$ keV, with 
unabsorbed luminosity $\approx 3 \times 10^{39}$ 
erg s$^{-1}$ in the $0.3$--$1$ keV band 
(Soria et al.~2004). It is still unclear 
whether this soft component, typical of many ULXs 
(Miller, Fabian \& Miller~2004; Feng \& Kaaret 2005), 
comes from the accretion disk or from 
a downscattering corona or outflow 
(e.g., Laming \& Titarchuk 2004; 
King \& Pounds 2003), or perhaps 
from a combination of reflection and absorption 
(Chevallier et al.~2006; Crummy et al.~2006); 
but in any case, it is difficult 
to reconcile with relativistically beamed emission.
Thus, it is more likely that at least the soft 
component of the X-ray emission 
is isotropic, requiring a BH with a mass 
$\ga 100 M_{\odot}$ to satisfy the Eddington limit. 
It is possible, instead, that the power-law component 
is beamed, for example produced by inverse-Compton 
scattering of disk photons by the jet 
(e.g., Bosch-Ramon, Romero \& Paredes 2005; 
Georganopoulos et al.~2002).

We conclude that Doppler boosting by a relativistic jet 
can in principle explain the observed radio flux 
and is not inconsistent with the X-ray properties. 
However, we consider this scenario somewhat contrived 
because it relies on the source spending most of its time 
in a ``persistent'' VHS with steep-spectrum dip-flare 
oscillations over many years. This is unlike the typical 
spectral state behaviour of Galactic X-ray 
binaries and microquasars; no other sources 
with the same behaviour have been found yet.
The other difficulty is that we have at least 
speculative evidence to believe that the radio source 
is extended over $\ga 30$ pc (Section 2), which 
would rule out point-like core emission.

\subsection{Supernova remnant?}

In all the scenarios discussed---and in most cases  
ruled out---so far, the radio emission would be coming from 
the innermost region around the BH, 
and would be directly related to its current 
state of accretion. 
We shall now consider the possibility 
that the emission comes, instead, 
from a larger region in the interstellar 
medium around the BH. 

The association with free-free radio emission 
from a star-forming region was already 
ruled out (Kaaret et al.~2003) 
due to the observed steep spectrum, 
inconsistent with thermal bremsstrahlung. 
A steep optically-thin synchrotron spectrum would 
instead be consistent with an underlying shell-like 
SNR: remnants in this class have typical radio spectral 
indices $-1 \la \alpha_{\rm R} \la -0.4$. 
On the other hand, we can rule out filled-center 
(plerionic, or pulsar-wind-nebula) 
SNRs---such as the Crab---because their spectral indices 
are much flatter ($-0.3 \la \alpha_{\rm R} \la -0.1$; 
e.g., Weiler \& Panagia 1978).
Our observed index $\alpha_{\rm R} \approx -1$ 
would be among the steepest, but not unique.
For example, at least three, and possibly five 
of the 14 shell-like SNR candidates in NGC\,300 
have a spectral index $\la -1$ (Pannuti et al.~2000); 
three of the 15 shell-like SNR candidates 
in NGC\,6946 have an index between $-1.4$ and $-1.0$ 
(Hyman et al.~2000); two of the 53 SNRs 
in M\,33 (including both plerion and shell-like SNRs) 
have an index $\approx -1.1$ (Duric et al.~1995).

The radio flux of our source in NGC\,5408 ($\approx 0.28$ 
mJy at 4.8 GHz) would also be among 
the highest for SNRs in nearby galaxies, 
but not unique. The brightest radio SNRs 
in NGC\,300 would only reach a flux density 
$\approx 0.06$ mJy at the distance of NGC\,5408.
However, 6 of the 15 radio SNR candidates in 
NGC\,6946 would have a 5 GHz flux $\ga 0.2$ mJy 
(Hyman et al.~2000); the brightest SNR in that galaxy 
would have a flux $\approx 0.5$ mJy.
All five of the candidate SNRs in NGC\,4258 
(Hyman et al.~2001) are also brighter, 
with fluxes $\approx 0.6$--$0.7$ mJy 
at the distance of X-1.
In NGC\,7793, four of the five brightest radio SNRs 
would have fluxes $\approx 0.08$--$0.12$ mJy 
at the distance of NGC\,5408 (Pannuti et al.~2002); however, 
the brightest candidate SNR in that galaxy, 
NGC\,7793-S26, has a spectral index 
$\alpha_{\rm R} = -0.9 \pm 0.2$ and would 
have a 5 GHz flux $\approx 0.6$ mJy.
The brightest shell-like SNR in our Galaxy, 
Cas A, would be seen with a 5 GHz flux 
of $\approx 0.3$ mJy (Reynoso \& Goss 2002) 
and a diameter of $\approx 0\farcs2$.

One explanation for such exceptionally 
bright SNRs is that they are produced 
by very massive stars with strong winds; 
hence, the explosion occurred in a dense 
circumstellar environment (Dunne, 
Gruendl \& Chu 2000). Alternatively, 
the remnants may be the result of multiple 
SN explosions, perhaps inside a star cluster 
(Pannuti et al.~2002). Either scenario 
would be consistent with a direct physical 
association between the SNR and the process 
of formation of the ULX X-1. In addition, 
a relatively young age may play a role.
In summary, a bright shell-like SNR 
would be consistent with the measured fluxes 
and spectral index, would explain 
why the flux has been approximately the same 
between 2000 and 2004, and would be 
a natural candidate for the formation 
of the ULX. 
If, as we speculate, the radio source in NGC\,5408 
is marginally resolved with a diameter $\sim 50$--$70$ pc, 
we can apply the purely empirical correlation between 
age and diameter, found in Galactic SNRs (Xu, 
Zhang \& Han 2005): we obtain an age $\ga 10^4$ yr, 
with a best-fit age $\approx 10^5$ yr (but with 
a large scatter). Thus, the source 
would be too large to be a very young 
or historical SNR (like Cas A).
In Sections 3.5 and 3.6 we shall 
try to find more stringent tests to the SNR 
scenario, based on the minimum-energy 
condition, and suggest an alternative explanation.

\begin{figure}
\epsfig{figure=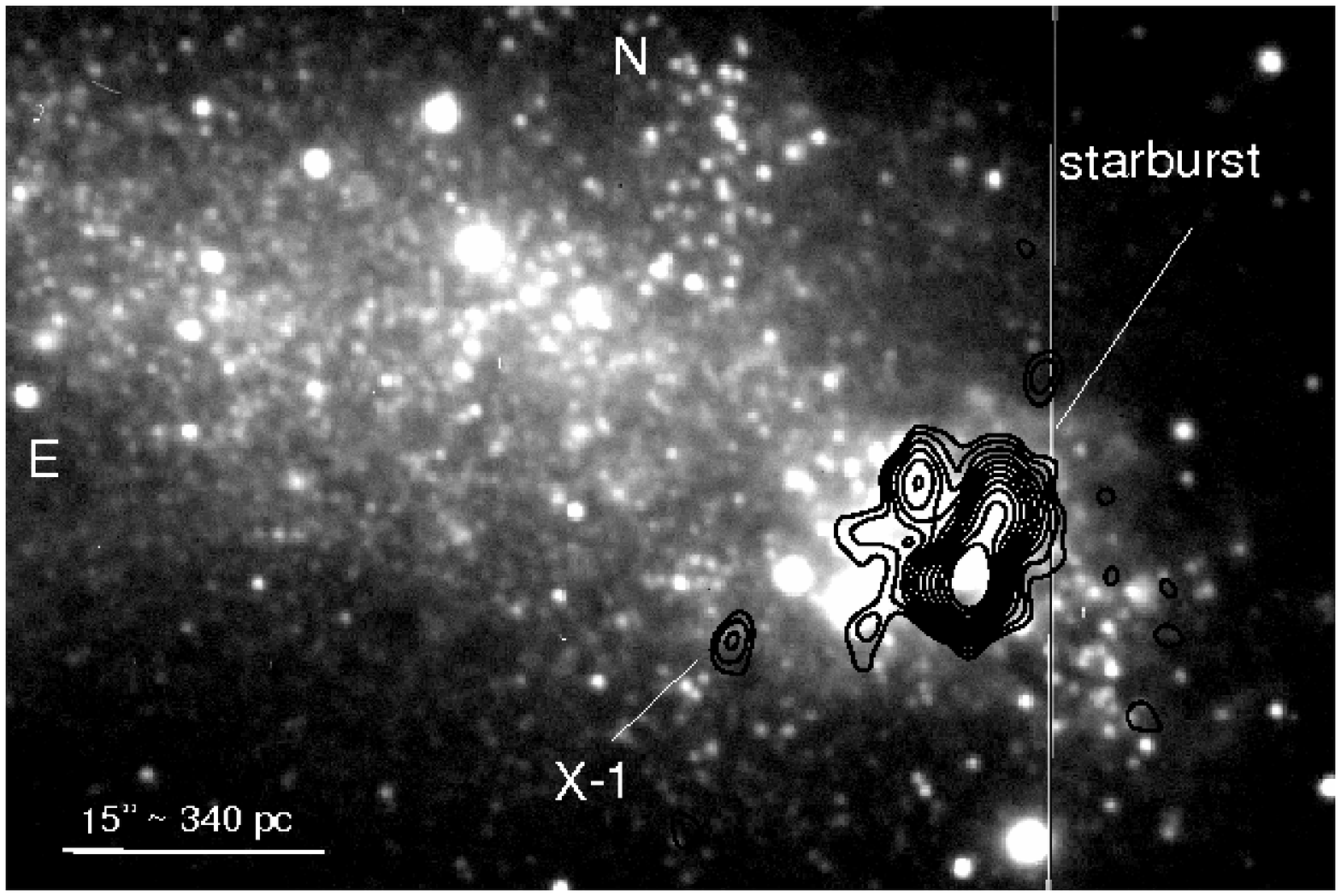, width=8.6cm}
\epsfig{figure=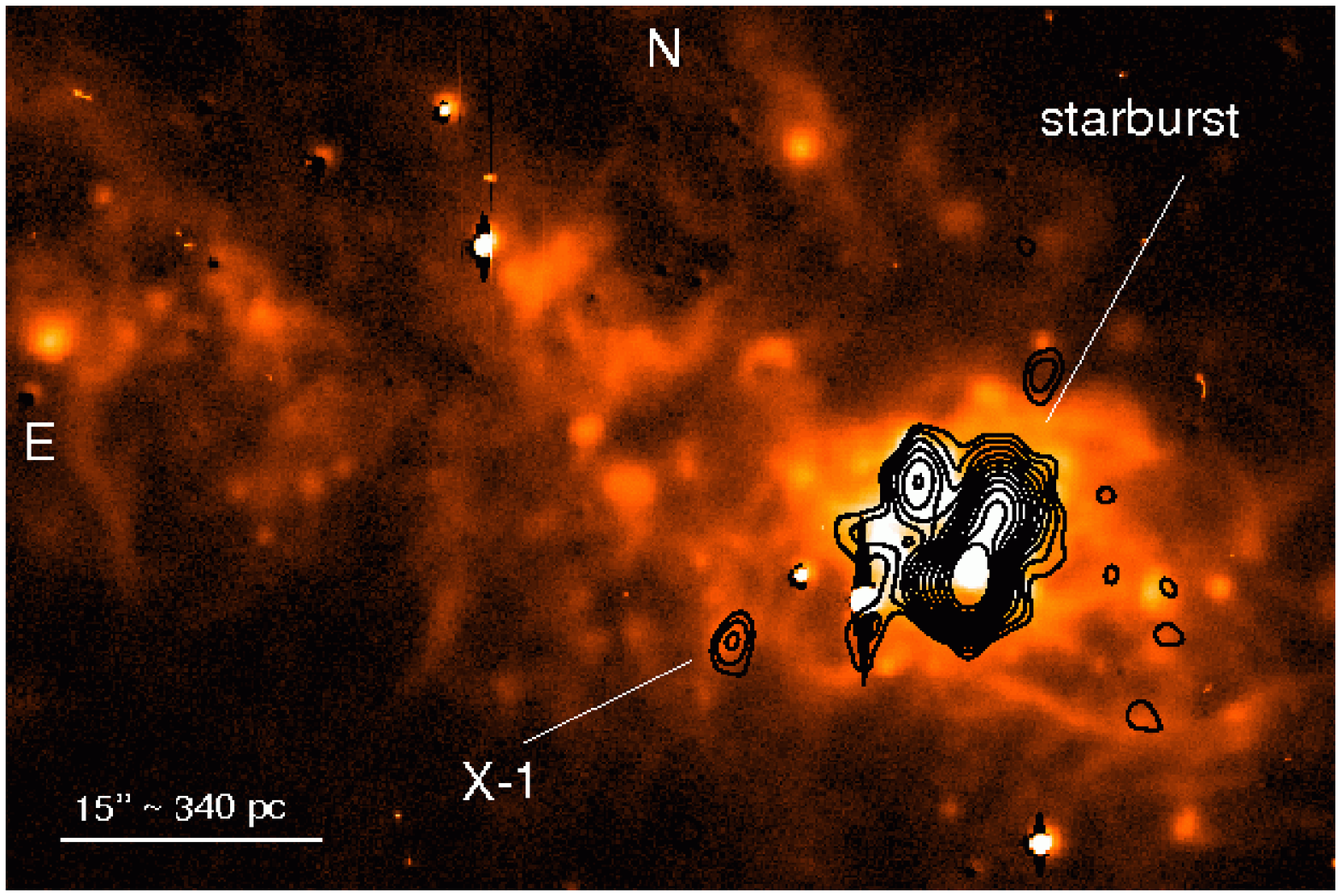, width=8.6cm}
\caption{ATCA radio flux density contours at 4.8 GHz, overplotted 
on a B image from {\it Subaru} (top panel) and an H$\alpha$ 
image from the ESO 3.6m telescope (bottom panel), 
on the same spatial scale.}
\end{figure}

\begin{figure}
\epsfig{figure=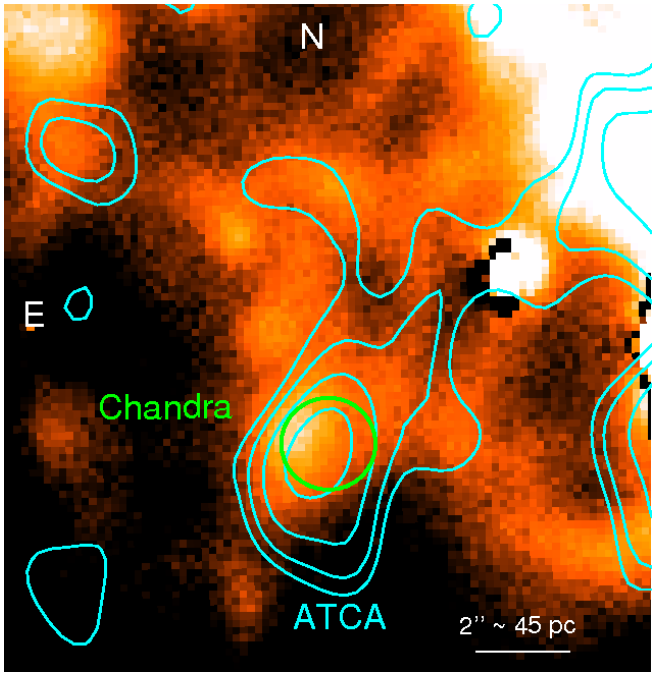, width=8.6cm}
\caption{Close-up view of the region around X-1, 
in the continuum-subtracted H$\alpha$ image from the ESO 
3.6-m telescope. A circle (green in the on-line version)
marks the {\it Chandra} position; its 1\arcsec radius 
combines the rms error in the astrometric solution 
for the optical image with the astrometric uncertainty 
in the {\it Chandra} position. ATCA 4.8-GHz radio flux 
density contours are overplotted (cyan, in the on-line version): 
the contours are 
for fluxes of 0.054, 0.081, 0.12 and 0.18 mJy beam$^{-1}$.}
\end{figure}

\subsection{H$\alpha$ emission: SNR or X-ray photoionization?}

The SNR scenario was initially rejected (Kaaret et al.~2003) 
because there appeared to be no evidence 
of optical emission lines at the X-1 position 
(upper limit of $\approx 7 \times 10^{-15}$ erg 
s$^{-1}$ cm$^{-2}$ derived for H$\alpha$ by 
Kaaret et al.~2003, from a broad-band {\it HST}/WFPC2 
image in the F606W filter).
Although the ULX appears to be well separated 
from the starburst region, in broad-band 
optical images (e.g., the $B$-band {\it Subaru} image 
in Figure 6, top panel; and the {\it HST}/WFPC2 image 
in Kaaret et al.~2003), a narrow-band H$\alpha$ 
image (Figure 6, bottom panel) shows instead that most 
of the galaxy is filled by a tangled web 
of H{\footnotesize{II}} filaments and knots.
For this study, we downloaded and analysed 
public-archive H$\alpha$ and R-continuum 
images taken from the ESO 3.6m telescope 
(EFOSC camera) by M. Pakull and L. Mirioni. 

In particular, an H$\alpha$-emitting knot 
is visible at the ULX position (Figure 7) 
(as noted in Mirioni 2003, Pakull \& Mirioni 2003):
it has a slightly elongated shape, with a size 
of $\approx 50 \times 70$ pc, matching the 
upper limit (and possibly the true size) 
of the radio-emitting blob. We checked that 
the slight displacement between the H$\alpha$ knot and 
the radio and X-ray positions is within the 
error in the astrometry of the optical 
image ($\sim 1\arcsec$); therefore, the H$\alpha$, 
radio and X-ray emission are consistent with being 
coincident. The H$\alpha$ knot is connected 
to the main starburst region through a network 
of filaments and cavities (Figure 7). 
We estimate a flux 
$f_{\rm {H}\alpha} = (5 \pm 2) \times 10^{-16}$
erg cm$^{-2}$ s$^{-1}$ from that knot, 
corresponding to a luminosity 
$L_{\rm {H}\alpha} = (1.5 \pm 0.4) \times 10^{36}$ 
erg s$^{-1}$ in the line (or $\approx 5 \times 
10^{47}$ H$\alpha$ photons s$^{-1}$), 
that is $\sim 10^{-3}$ times 
the total H$\alpha$ emission from NGC\,5408. 
The peak surface brightness  
$S_{\rm {H}\alpha} \approx 1.5 \times 10^{-16}$
erg cm$^{-2}$ s$^{-1}$ arcsec$^{-2}$.

If the H$\alpha$ line emission is due to recombination, 
its estimated luminosity requires 
a flux of $\sim 1$--$2 \times 10^{48}$ ionizing 
photons s$^{-1}$ intercepted by the nebula 
(Osterbrok 1989; cf.~also similar calculations 
for other photo-ionized nebulae in Pakull \& Angebault 1986, 
and Soria et al.~2005).
This is a relatively modest requirement.
Continuum {\it Subaru} and {\it HST} images show three massive 
stars near the peak of the radio and H$\alpha$ emission, 
with absolute $B$ magnitudes $\approx -6$ and 
bluish colors, $B-V \la 0$
(Kaaret et al.~2003; Copperwheat et al.~2006, in prep.).
They are consistent with main-sequence O stars, 
or, more likely, blue supergiants.
Three such stars with initial masses $\sim 20 M_{\odot}$ 
would provide enough ionizing photons 
to explain the H$\alpha$ emission 
(Sternberg, Hoffmann \& Pauldrach 2003; 
Schaerer \& de Koter 1997).
As a consistency check, the size 
of a Str\"{o}mgren sphere at a characteristic 
temperature $T \approx 10^4$ K is  
$R_{\rm S} \approx 30\, Q_{48}^{1/3} n_{\rm H}^{-2/3}$ pc, 
where $Q_{48}$ is the photo-ionizing flux 
in units of $10^{48}$ photons s$^{-1}$ 
and $n_{\rm H}$ is the hydrogen number density in units 
of $1$ cm$^{-3}$. This is in agreement with 
the observed size of the H\,{\footnotesize {II}} region
at the ULX position\footnote{Note that 
the thermal (free-free) radio emission  
associated with a photoionized H\,{\footnotesize {II}} 
region of such size and luminosity
is $\approx 500$ times fainter 
than the non-thermal emission measured 
at the ULX position (Dickel, Harten \& Gull 1983; 
Lazendic, Dickel \& Jones 2003). Hence,  
the thermal contribution is absolutely negligible 
in our case, and does not affect our estimate 
of the radio spectral index.}.

We stress that this H$\alpha$ nebula 
is not at all unique or remarkable for this galaxy, 
and is fainter than many other knots and filaments 
(Figure 6), which are probably due to OB photo-ionization or  
local density enhancements in the outflowing 
warm gas. It is only its association with 
strong X-ray and synchrotron radio emission 
that makes it remarkable. Hence, we also need to consider 
the possibility that the local enhancement 
in the H$\alpha$ emission is directly caused by the ULX. 
We used {\footnotesize XSPEC} to re-analyze 
the {\it XMM-Newton} X-ray spectra 
of the ULX from 2001--2003 (Soria et al.~2004), 
using variable-abundance, ionized absorbers 
in addition to the Galactic line-of-sight term.
We found that intrinsic absorption (column 
density $\approx 1 \times 10^{21}$ cm$^{-2}$)
removes $\approx 1.7 \times 10^{48}$ 
photons s$^{-1}$ with energies above $13.6$ eV, 
assuming a thermal Comptonization model ({\tt bmc}) 
for the X-ray spectrum\footnote{Power-law plus blackbody 
models also provide good fits to the X-ray spectrum, 
but are less suitable for an extrapolation 
to the UV region, since they give unphysically 
high fluxes, unless the power-law component 
is truncated at low energies.}.
Plausible local interstellar gas densities 
$n_{\rm H} \sim$ a few cm$^{-3}$ 
are consistent with a significant fraction 
of the intrinsic absorption coming  
from $\la 30$ pc from the ULX.
Hence, photoionization from the ULX 
may contribute a similar amount of H-ionizing 
photons as the few OB stars in the surrounding 
region. 
 
Finally, part of the H$\alpha$ emission 
may be due to shock-ionization from an SNR. 
Most identified radio SNRs in nearby galaxies 
are associated with much brighter H$\alpha$ emission, 
from $\sim$ a few $10^{-15}$ erg s$^{-1}$ cm$^{-2}$ 
up to $\sim$ a few $10^{-14}$ erg s$^{-1}$ cm$^{-2}$ 
at the distance of X-1 (Hyman et al.~2000; 
Pannuti et al.~2000; Deharveng et al.~1988).
This is partly due to a selection bias
(e.g., Pannuti et al.~2000):
unresolved, non-thermal (steep-spectrum) radio 
sources without optical line emission would 
generally be discarded as background AGN.
However, radio synchrotron and H$\alpha$ emission  
in SNRs come from different processes, and are not strongly 
correlated; in fact, there are a few 
radio-bright candidate SNR in nearby galaxies, 
with H$\alpha$ emission below the level inferred for the 
source in NGC\,5408: for example, some sources in the Large 
Magellanic Cloud (LMC) (Figure 1a in Filipovic et al.~1998) 
\footnote{For comparison with the plot in Filipovic et al.~1998, 
the H$\alpha$ flux and 4.8 GHz radio flux 
density from the source in NGC\,5408 would be 
$\approx 5 \times 10^{-12}$ erg cm$^{-2}$ s$^{-1}$ 
and $\approx 2.5$ Jy, respectively,  
translated to the distance of the Large Magellanic Cloud.}.
Other examples of H$\alpha$-faint SNRs are found in NGC\,300 
(Table 4 in Payne et al.~2004).

In summary, we conclude that the SNR scenario 
for the radio synchrotron emission remains viable, 
although probably not the most likely. 
We shall also see in Section 3.6 that the radio source would require 
an explosion energy unusually high for a normal SN.
In the optical band, more narrow-band imaging and spectral studies 
are needed to distinguish between the main competing scenarios 
(H\,{\footnotesize{II}} region, X-ray ionized nebula, shock-ionized 
SNR). Specifically, the most important constraints 
will come from the presence or absence 
of nebular [O\,{\footnotesize{III}}] 
and He\,{\footnotesize{II}} emission, and from the 
[S\,{\footnotesize{II}}]/H$\alpha$ 
and [O\,{\footnotesize{I}}]/H$\alpha$ line ratios. 


\subsection{Radio lobes?}

Finally, we consider the possibility that 
the optically-thin synchrotron radio emission 
comes from extended radio lobes associated with the ULX.
While core emission from a radio jet (Sections 3.1--3.3) 
traces the instantaneous accretion power and spectral 
state of the BH, emission from radio lobes 
depends on the accretion activity integrated 
over the source lifetime (e.g., Scheuer 1974), 
even if the core were not currently active.
Hence, we do no expect significant variability 
over a few years, in agreement with the observations.

Almost all radio-loud AGN show a lobe structure, 
which dominates the radio emission in FR\,II sources.
On the other hand, while most X-ray binaries are 
thought to release part of their accretion power 
in a jet, only a few sources have been found so far 
with persistent lobe-like emission: Cyg X-1 (Gallo 
et al 2005); Cir X-1 (Tudose et al.~2006); the Galactic 
microquasars 1E 1740.7$-$2942 and GRS 1758$-$258
(Mirabel et al.~1992; Hardcastle 2005). 
In addition, in SS\,433, a radio-lobe structure 
may coexist with an underlying SNR (Downes, Pauls \& 
Salter 1996; Dubner et al.~1998); it has been 
noted that SS\,433 may be classified as a ULX if 
we could see it face-on (Fabrika \& Mescheryakov 2001).
The size of the steep-spectrum radio-lobe structure 
in SS\,433 is $\approx (50 \times 100) (d/3{\rm ~kpc})$ pc, 
very similar to the source in NGC\,5408; however, 
the radio luminosity density is an order of magnitude lower.

The main difference between radio galaxy and microquasar 
lobes is due to the environment:
while some quantities, such as all characteristic 
time-scales, length-scales and energies, 
scale as $M_{\rm BH}$, others, such as the density 
of the ambient medium, do not (e.g., Heinz 2002). 
In particular, radio lobes in stellar-mass microquasars 
tend to be much more over-pressured with respect 
to the external medium; therefore, they have a shorter 
lifetime due to adiabatic expansion (Hardcastle 2005; 
Hardcastle \& Worrall 2000). Nonetheless, 
Pakull \& Mirioni (2002) discovered that 
many ULXs are indeed surrounded by large optical nebulae 
possibly powered by the interaction of a jet 
with the interstellar medium.

The spectral index and total radio power alone 
do not give us enough information to determine 
the physical parameters of the lobes---in particular, 
the energy density of the particles and the magnetic 
field. However, if the size of the lobes (hence, 
their surface brightness) is known, one 
can minimize the total energy density 
to get a rough constraint on the particle 
and field energy and pressure (e.g., Pacholczyk 1970).
In our case, the radio source may be marginally resolved 
(Section 2), although its detailed spatial geometry is unknown. 
For simplicity, we take a spherical lobe with a radius 
of $30$ pc ($\approx 1\farcs3$), 
a power-law distribution of relativistic 
electrons with spectral index $-3$ (corresponding 
to $\alpha_{\rm R} = - 1$), and we assume that the electron 
distribution extends between Lorentz factors of $1$ 
and $10^4$~\footnote{For a random magnetic field $\approx 0.1$ mGauss, 
the synchrotron cooling timescale of relativistic electrons 
is $\approx 10^5 \times (10^4/\gamma)$ yr, where $\gamma$ 
is the Lorentz factor. Taking instead an upper limit 
$\gamma \sim 10^3$ for the electron distribution 
does not significantly affect our results: for our rather 
steep spectral index $\approx -1$, the minimum energy depends 
more strongly on the low-energy cut-off of the electron 
distribution. For example, a lower cut-off 
at $\gamma \sim 10$ would reduce the estimated energies 
by a factor of 3.}. We also assume a magnetic field randomly 
distributed with respect to the line of sight, and a volume 
filling factor $\approx 1$. See Appendix A for a brief 
summary of the formulae we used to estimate the minimum-energy 
magnetic field.

The matter content of a quasar or microquasar jet 
is still a hotly debated issue (Romero 2005). 
For example, the Galactic microquasar LS 5039 
has been variously modelled with leptonic 
(Bosch-Ramon et al.~2005; 
Dermer \& B\"ottcher 2006) or hadronic jets
(Romero et al.~2003; Aharonian et al.~2005). 
An electron-positron plasma was suggested  
for the jet in Nova Muscae (GRS\,1124$-$684) 
(Kaiser \& Hannikainen 2002). On the other hand, 
a significant hadronic contribution could explain 
why synchrotron emission is not detected within 
the free-free emitting bubble of Cyg X-1 (Gallo et a.~2005).
Hadronic jet models are also favoured 
for the microquasars SS 433 (Migliari, Fender \& Mendez 2002), 
and LS I\,$+$61\,303 (Romero, Christiansen \& Orellana 2005).
A leptonic model was instead suggested for the powerful quasar
3C 279 (Wardle et al.~1998), based on the detection of circular 
polarization.

From the observed 4.8 GHz flux, 
for an electron-positron plasma, 
we obtain: a minimum energy density $\approx 9.4 \times 10^{-10}$ 
erg cm$^{-3}$ for a random magnetic field 
$B \approx 0.11$ mGauss\footnote{For the adopted spectral index 
$\alpha_{\rm R} = - 1$, the minimum-energy condition 
also implies energy equipartition, while for a spectral index $= -0.7$ 
the energy density of the magnetic field would be $\approx 85\%$ 
of the particle energy density.}; 
a minimum pressure 
$P_{\rm in} \approx 5.4 \times 10^{-10}$ dynes cm$^{-2}$; 
and a total energy $\approx 3.3 \times 10^{51}$ erg. 
Alternatively, taking two lobes with radii 
of $23$ pc $= 1\arcsec$,  
the minimum pressure is $\approx 6.2 \times 10^{-10}$ dynes cm$^{-2}$
and the total energy is $\approx 2 \times 1.7 \times 10^{51}$ erg.
Most quantities depend strongly on the adopted value 
of $\alpha_{\rm R}$. Taking $\alpha_{\rm R} = -0.7$ 
as a strong upper limit, we infer 
a total energy $> 3.5 \times 10^{50}$ erg 
for a single $30$-pc electron-positron bubble, 
or $> 3.6 \times 10^{50}$ erg for two $23$-pc lobes.

If energy and pressure in the synchrotron-emitting bubble 
are dominated by relativistic protons, 
the total energy can be up to a factor of 10 higher than 
in the leptonic case, for the same sets 
of parameters. For example, if hadrons carry $\sim 99\%$ 
of the energy\footnote{a value typical of plasmas where 
electrons and positrons originate 
as secondary particles following collisions of a primary 
hadronic jet with the ISM (Pacholczyk 1970).}, 
the minimum-energy condition implies 
a magnetic field $B \approx 0.35$ mGauss and a total 
energy $\approx 3.3 \times 10^{52}$ erg for a $30$-pc 
synchrotron bubble with a spectral index $\alpha_{\rm R} = -1$; 
our limit $\alpha_{\rm R} < -0.7$ implies an energy 
$> 4.3 \times 10^{51}$ erg.
Hence, the leptonic-jet assumption can be used 
as a conservative lower limit to the energy in the bubble.

On the other hand, relativistic hadrons do carry 
most of the energy in SNRs (known sources of cosmic rays).
Therefore, if the radio source in NGC\,5408 is interpreted 
as an SNR, we need to use the higher range 
of energy estimates. Energies $\sim 10^{52}$ erg
are implausibly high for a single ``standard'' SN. 
Following this argument leads us to conclude that   
a SNR interpretation is less likely than a radio lobe 
scenario, but we cannot rule out the possibility of  
a more energetic hypernova remnant.
Conversely, we could assume that the source is an SNR 
with a standard energy $\approx 10^{51}$ erg, 
and infer its size from the minimum-energy condition.
It turns out that the remnant should be $\approx 10$--$15$ pc 
in radius, for the observed flux and spectral index. 
Hence, deeper high-frequency radio observations 
will be crucial to unequivocally determine  
the size of the radio source (or at least 
a more stringent upper limit) and test the SNR scenario.

One might suggest that the injection spectral index 
was flatter ($\alpha_{\rm R} \approx -0.5$), and  
the steeper index currently observed at high frequencies 
may simply be due to an ageing population of relativistic 
electrons inside the expanding bubble (``spectral ageing''), 
as the higher-energy electron population is depleted 
by synchrotron losses. If this is the case, 
the total energy in a $30$-pc bubble would be
lower, $\approx 1 \times 10^{51}$ erg  
(corresponding to a magnetic field $B \approx 0.07$ mGauss), 
consistent with a normal SN. 
The presence of a spectral break in the radio continuum 
would, in principle, provide another criterion 
to distinguish between an SNR and a jet-inflated lobe.
For an SNR, we expect an exponential cut-off 
in the spectrum above the break frequency $\nu_{\rm br}$
(Jaffe \& Perola 1973); for a jet bubble, 
characterized by continuous injection, 
the spectrum would be a broken power law 
with an index $\alpha_{\rm R}' = \alpha_{\rm R} - 0.5$ 
above the break frequency (Kardashev 1962).

The problem with the spectral break scenario is that 
a relatively old age is required for the break 
to occur at frequencies as low as $\sim 3$--$4$ GHz. 
Following Murgia (2003), and using the previous 
order-of-magnitude estimate of the magnetic field, 
we obtain a characteristic age 
$\approx 8 \times 10^{5} \, (B/0.1{\rm ~mG})^{-1.5} (\nu_{\rm br}/4 {\rm ~GHz})^{-0.5}$ yr 
for an expanding $30$-pc bubble. SNRs of that age 
are unlikely to be still so bright.
This age is also longer than the characteristic 
age we shall estimate for the radio-lobe scenario 
(Section 4).
However, even in this case we cannot entirely rule 
the break scenario, for example if $B$ is well above 
equipartition in a magnetically dominated outflow.

\section{Self-similar solution for radio lobes}

In Section 3.6 we have treated the size of the bubble 
simply as an arbitrary input parameter.
We can learn more about the time evolution of a jet-inflated bubble 
if we express its radius self-consistently 
as a function of the total energy injected into it, 
thus coupling a radial expansion equation with the energy density 
equation. To do so, we describe the supersonic (but sub-relativistic) 
expansion and radio emission from the bubble with a set 
of self-similar solutions (Castor, McCray \& Weaver~1975; 
Falle 1991), often adopted to model radio lobes in AGN and Galactic 
micro-quasars (Heinz, Reynolds \& Begelman 1998; Willott et al.~1999; 
Jarvis et al.~2001; Heinz 2002). We assume (Heinz et al.~1998): 
a spherical, uniform radio cocoon of radius $r_{\rm c}$; 
a constant density of the external ISM (an approximation suitable 
to microquasar environments); relativistic gas inside 
the radio bubble (adiabatic index $= 4/3$); non-relativistic 
gas in the shell outside the contact discontinuity 
(adiabatic index $=5/3$); equipartition of field 
and particle energies in the cocoon; non-radiative shocks; 
a radio spectral index $\alpha_{\rm R} = -1$. For simplicity, 
we use a density $\rho = 1.7 \times 10^{-27} n_{\rm H}$, 
valid for a pure hydrogen gas; in any case, the main results 
from the self-similar solutions do not have 
a strong dependence on the mean atomic weight.

The self-similar equations governing the expansion 
of the bubble are (e.g., Heinz et al.~1998):
\begin{equation}
r_{\rm c} \simeq 4.02 \times 10^4 n_{\rm H}^{-1/5} {\mathcal L}^{1/5} t^{3/5},
\end{equation}
\begin{equation}
{\mathcal L} t \equiv E_{\rm tot},
\end{equation}
where $E_{\rm tot}$ is the total energy 
injected by the jet into the radio cocoon,  
and ${\mathcal L}$ is the time-averaged jet power, 
assuming that it is either approximately constant, 
or with a duty cycle much shorter than the time $t$.
The emitted flux density at a frequency $\nu$ is then
\begin{eqnarray}
F_{\nu} &=& 0.41 \left(\frac{n_{\rm H}}{{\rm~cm}^{-3}}\right)^{0.6} 
\left(\frac{{\mathcal L}}{10^{39} {\rm ~erg~s}^{-1}}\right)^{1.4} 
\left(\frac{t}{10^5 {\rm ~yr}}\right)^{0.2} \nonumber\\
&&\times \left(\frac{\nu}{5 {\rm ~GHz}}\right)^{-1} 
\left(\frac{d}{4.8 {\rm ~Mpc}}\right)^{-2} \ {\rm mJy}.
\end{eqnarray}
See Appendix B for an outline of the derivation of Eq.(5).

We followed Willott et al.~(1999) to obtain the normalization 
of the relations between jet power and age and radio flux 
of the cocoon (Equations 3 and 5)\footnote{A similar  
method was followed by Heinz (2002), see his Eq.~(3), but 
with the assumption of a radio spectral index $\alpha_{\rm R} = -0.5$, 
definitely ruled out in our case.}.
The normalization factors are based on the estimated 
ages, ambient gas densities and $151$-MHz flux densities 
for the FR\,II radio galaxies in the 7C Redshift survey.
They also depend on the adopted value of a scaling factor $f$ 
(as defined in Willott et al.~1999) for the total energy 
inside the cocoon, inferred from the minimum-energy 
condition. The factor $f$ parameterizes our ignorance 
of the exact geometry 
of the cocoon, low-energy cut-off in the electron distribution, 
filling factor, deviations from equipartition, and 
proton contribution. The assumptions we adopted in Section 3.6 
to estimate the total energy correspond to a choice 
of $f\approx 15$, which is a plausible number for QSO 
radio lobes; it was estimated in Blundell \& Rawlings (2000) that 
$f$ is most likely to vary between $\sim 10$ and $20$.
Therefore, our normalization coincides with the one adopted by 
Punsly (2005), that is ${\mathcal L} = 4.0 \times 10^{46}$ 
erg s$^{-1}$ for $d^2 F_{151} = 10^{35}$ erg s$^{-1}$ Hz$^{-1}$ 
sr$^{-1}$.
This choice of normalization may easily vary by a factor
of 2 at the reference flux, and larger uncertainties 
may be introduced as we scale the solution from the parameter 
range suitable for radio-loud QSOs (e.g., $n_{\rm H} \sim 10^{-3}$ cm$^{-3}$, 
$t \sim 10^7$ yr, ${\mathcal L} \sim 10^{46}$ 
erg s$^{-1}$), to the range suitable for microquasars 
(e.g., $n_{\rm H} \sim 1$ cm$^{-3}$, 
$t \sim 10^5$ yr, ${\mathcal L} \la 10^{39}$ 
erg s$^{-1}$). Another caveat is that sources with 
a radio spectral index $\approx -1$ 
are outliers in the FR\,II sample: 
most have indices between $\approx -0.5$ and $\approx -0.9$ 
(e.g., Figure 5 in Willott et al.~1999).

Inserting the observed flux ($0.28$ mJy at $4.85$ GHz) 
and speculative size ($r_{\rm c} \approx 30$ pc) 
of the radio-emitting bubble into Equations (3--5), 
for an electron-positron plasma,
we obtain an age $t \approx 1.43 \times 10^5 n_{\rm H}^{1/2}$ yr, 
an average jet power ${\mathcal L} = 7.1 \times 10^{38} n_{\rm H}^{-1/2}$ 
erg s$^{-1}$, a total energy $E_{\rm tot} \approx 3.2 \times 10^{51}$ 
erg, and an expansion velocity 
$v \approx (2r_{\rm c}/5t) \approx 82 n_{\rm H}^{-1/2}$ km s$^{-1}$ 
at present time. 
We recall that we estimated (Section 3.6) a total pressure  
$P_{\rm in} \approx 5 \times 10^{-10}$ dynes cm$^{-2}$ 
inside the lobes; this is much larger than 
the thermal pressure of the undisturbed ISM, which is   
$P_{\rm out} \approx 3 \times 10^{-12} (n_{\rm H}/1{\rm ~cm}^{-3}) (T/10^4 {\rm~K})$ 
dynes cm$^{-2}$, with a sound speed $\approx 15$ km s$^{-1}$ 
for $T \approx 10^4$ K.
The expansion of the bubble into the ISM is supersonic, 
with Mach number $\mathcal M \approx 82/15 \approx 5.5$. 
Hence, we expect the expanding cocoon to drive a strong shock 
into the ISM. Pressure balance should occur 
between the pressure inside the cocoon ($P_{\rm in}$) 
and the thermal plus ram pressure of the shocked ambient gas, 
swept up in a shell outside the cocoon.
Applying standard Rankine-Hugoniot conditions 
(e.g., Spitzer~1978), we expect a thermal pressure 
in the shocked gas 
$P_{\rm s} = (5{\mathcal M}^2 -1)P_{\rm out}/4  \approx 37 n_{\rm H}^{-1} P_{\rm out} 
\approx 1 \times 10^{-10}$ dynes cm$^{-2}$, 
a density of the shocked gas $\rho_{\rm s} \approx 3.6$ times the 
ambient ISM density, and a ram pressure 
$\approx \rho_{\rm s} v^2 \approx 4 \times 10^{-10}$ dynes cm$^{-2}$.
Thus, even though our simple spherical model is only 
a crude approximation of a real radio lobe, 
we obtain that the total pressure 
inside the cocoon is balanced 
by the thermal plus ram pressure of 
the swept-up interstellar medium, 
as we should expect. 

A higher jet power and shorter bubble age are inferred if protons 
contribute a significant fraction of energy. 
For example, in the case of a $99\%$ hadronic 
contribution, we obtain an age $t \approx 5 \times 10^4 n_{\rm H}^{1/2}$ yr, 
an average jet power ${\mathcal L} = 2 \times 10^{40} n_{\rm H}^{-1/2}$ 
erg s$^{-1}$, and an expansion velocity 
$v \approx 230 n_{\rm H}^{-1/2}$ km s$^{-1}$ 
at present time. The total energies and pressure inside 
the bubble are an order of magnitude higher 
than in the leptonic case ($\sim B^2$). 
The internal pressure is still roughly balanced by ram plus thermal 
pressure in the shocked gas ($\sim v^2$).

Finally, from the Rankine-Hugoniot equations we also 
predict a temperature in the swept-out shell 
$\approx 10$ times the ambient ISM temperature, 
i.e. $\approx 10^5$ K. The shell thickness should be 
$\approx 0.1 r_{\rm c} \approx 3$ pc. We infer a cooling 
timescale for the shocked gas $\approx 10^6$ yr, that is, 
longer than the estimated lifetime of the radio cocoon. 
This may explain why we do not see stronger Balmer 
line emission from that region. The total synchrotron 
energy radiated over the lifetime of the bubble 
is also negligible ($O(10^{-2})$) compared 
with the injected energy ${\mathcal L} t$ (Willott et al.~1999).
This explains why the value of $E_{\rm tot}$ 
in Equation (4) can be treated as the total energy 
inside the bubble at present time.

Considering the various sources of uncertainty 
mentioned earlier, the scenario outlined 
in this Section can only represent 
one possible solution in agreement with 
the (few) observed parameters. Nonetheless, 
it is a plausible scenario, consistent 
with existing models for microquasars.
As already mentioned in Section 3.1, a steady 
radio jet is thought to dominate the power output in the LHS, 
and a transient jet may be present in the VHS, 
while it is suppressed in the HS.
In the conventional definition of LHS, 
the X-ray luminosity $L_{\rm X} \la 10^{-2} L_{\rm Edd}$, 
and the jet power
$L_{\rm J}/L_{\rm Edd} \approx A (L_{\rm X}/L_{\rm Edd})^{1/2}$, 
with $6\times 10^{-3} \la A \la 0.3$ (Fender, Gallo \& Jonker 2003;
Fender et al.~2004; Malzac, Merloni \& Fabian 2004). Hence, 
the jet power in the LHS is $L_{\rm J} \la 0.03 L_{\rm Edd}$. 
We compare this value with the value 
of ${\mathcal L} \approx 7 \times 10^{38}$ erg s$^{-1}$ 
inferred from our self-similar solutions.
If most of the energy has been injected into the bubble 
by a steady LHS jet over the past $\sim 10^5$ yr, 
the Eddington luminosity has to be $\ga 2 \times 10^{40}$ 
erg s$^{-1}$, corresponding to a BH mass $\ga 150 M_{\odot}$.
However, the upper limit to the luminosity 
in the LHS is somewhat conventional: a steady jet 
may persist (with a similar scaling) 
as the source moves up in luminosity 
into the ``hard'' VHS, at 
$L_{\rm X} \approx L_{\rm Edd}$ (Fender et al.~2004).
If the source spent most of its time 
in that state, $L_{\rm Edd} \sim$ a few $10^{39}$ erg s$^{-1}$ 
would suffice. Finally, transient ejections with 
$L_{\rm J} \sim 0.1$--$1 L_{\rm Edd}$
occur at the spectral transition between ``hard'' and ``soft'' VHS.

We do not know how much time the X-ray source has spent 
in the different accretion states, 
and for what fraction of time it has been off. 
As we discussed in Section 3 (see also Soria et al.~2004), 
it is currently in some kind of VHS, with 
$L_{\rm X} \approx 10^{40}$ erg s$^{-1}$ and some 
ongoing flaring activity suggesting coronal ejections.
The estimated average jet power is at least 
consistent with a BH mass $\approx 100 M_{\odot}$.
For comparison, GRS 1915$+$105 ($M_{\rm BH} \approx 14 M_{\odot}$) 
spends long periods of time in a flaring state with 
$L_{\rm J} \approx 10^{38}$ erg s$^{-1}$ 
and $L_{\rm X} \approx 10^{39}$ erg s$^{-1}$
(Fender et al.~1999).
The average jet power in SS\,433 is thought to be 
$\sim 10^{39}$ erg s$^{-1}$ (Fabrika et al.~2004, and 
references therein). In the ULX class, 
a jet power $\approx 1.5 \times 10^{39}$ erg s$^{-1}$ 
$\approx 0.2 L_{\rm X}$ would 
be required to explain the ionized bubble around 
NGC\,1313 X-2 (Pakull \& Mirioni 2002).
Another example---at the opposite end of the BH mass range---is 
the quasar PKS\,0743$-$67, showing a radio jet 
with $L_{\rm J} \sim 0.1 L_{\rm Edd} \sim 10^{46}$ erg s$^{-1}$ 
as well as an accretion luminosity $\sim L_{\rm Edd}$ 
(Punsly \& Tingay 2005).

\section{Conclusions}

We have used the ATCA to study the radio counterpart 
of a bright ULX in the starburst irregular 
galaxy NGC\,5408. The spectrum is rather steep 
(single power-law index $\approx -1$), typical 
of optically-thin synchrotron. 
The flux has remained unchanged between March 2000  
and December 2004. We speculate that the source is 
marginally resolved in a composite (4.8 + 6.2) GHz map, 
implying a size $\approx 50 \times 70$ pc.  
However, deeper 6.2 or 8.4 GHz observations will 
be necessary to verify this hypothesis unequivocally, 
and to measure the spectral index more precisely.

We have discussed the possibility 
that the source represents the core radio emission 
from a ULX jet, as previously hypothesized 
(Kaaret et al.~2003). We argued that a steady jet 
in the LHS is inconsistent with its steep radio spectrum, 
its very high flux ($\approx 0.28$ mJy at 4.8 GHz), 
and the X-ray spectral and timing properties of the ULX.
Therefore, simple BH mass determinations 
based on fundamental-plane relations 
between LHS core radio and X-ray luminosities   
are not applicable to this object.
A sporadic, major radio outburst is also ruled out, given 
the constant radio flux between the two epochs of 
the ATCA observations (2000 and 2004), 
and the X-ray behaviour. Radio emission from 
a persistent, flaring VHS
(perhaps similar to the dip-flare state in GRS 1915$+$105, 
characterized by a sequence of repeated, minor ejections)
can be consistent with the observed flux, if enhanced by 
(moderate) relativistic Doppler boosting.
We consider this scenario slightly contrived, 
because there are currently no known examples 
of similar, persistent VHS flaring in Galactic 
microquasars. On the other hand, if the accreting 
BH has a mass $\sim 100 M_{\odot}$ as suggested 
by the X-ray luminosity (from the Eddington argument), 
the spectral state transition pattern may be rather 
different from Galactic microquasars.
The other reason why we consider core radio emission 
an unlikely scenario is that we have at least 
speculative evidence of a spatially resolved source. 

Instead, we suggest that the emission 
could be either from an underlying SNR 
(perhaps related to the formation 
of the compact object powering the ULX), 
or from radio lobes inflated by a relativistic jet from the ULX.
A young shell-like radio SNR could reach sufficient 
luminosity, would have a constant flux over many years, and 
has an optically-thin synchrotron spectrum.
However, in contrast with its unusually high radio luminosity, 
this source is extremely faint in the optical.
From a continuum-subtracted H$\alpha$ image, 
one can see an H\,{\footnotesize{II}} 
region approximately coincident with the X-ray 
and radio sources. This region appears as a bright knot 
in a complex web of H$\alpha$-emitting filaments, 
connected to the main starburst complex.
We estimate a luminosity $\approx 1.5 \times 10^{36}$ 
erg s$^{-1}$ in the line. This is an upper limit 
to any possible contribution from an SNR. In fact, 
we showed that the few OB stars clearly detected 
in the same region, and/or the X-ray source 
itself would contribute enough ionizing photons 
to explain the H$\alpha$ emission.
A study of higher-ionization lines will  
be required to determine the ionization mechanism.
Moreover, if we accept the evidence for a marginally-resolved 
radio source, standard equipartition arguments 
imply a total energy $\ga$ a few $\times 10^{51}$ erg, 
and probably as high as $\sim 10^{52}$ erg 
in the SNR: this is extremely high 
for a single SN, and may require a hypernova or 
a few normal SNe going off in the same group 
of stars. We conclude that the SNR interpretation 
is unlikely, though not ruled out.

A radio lobe powered by a jet 
is thus the most likely scenario, in our opinion.
In equipartition, the total energy content 
is $\sim 10^{51}$ erg for a leptonic jet, 
or up to an order of magnitude higher if there is 
a significant hadronic contribution.
We modelled the lobe as a spherical 
bubble of radius $30$ pc, and solved a set of  
self-similar equations---often used 
in the literature to model radio lobes in FR\,II 
quasars---to relate the bubble expansion 
with the energy input (i.e., the jet power) 
and the observed radio flux and spectral index. 
Assuming an electron-positron plasma, we obtain 
that an average jet power $\approx 7 \times 10^{38}$ 
erg s$^{-1}$ over a time-scale $\approx 1.4 \times 10^5$ yr 
would indeed produce a $30$-pc cocoon with the observed 
radio brightness, a total energy content 
$\approx 3 \times 10^{51}$ erg, with a randomly-oriented 
magnetic field $\approx 0.1$ mGauss.
Such a bubble expands into the ambient ISM 
with a present speed of $\approx 80$ km s$^{-1}$, 
sweeping up a shell of shocked gas. The pressure 
inside the bubble ($\approx 5 \times 10^{-10}$ dynes cm$^{-2}$) 
is balanced by the thermal plus ram pressure of the 
gas in the swept-up shell.
An age $\sim 10^5$ yr implies that the steepness 
of the spectrum around 5 GHz is from injection 
and cannot be due to ageing (unless the magnetic field 
is well above equipartition); 
the age is also shorter than the radiative cooling 
timescale of the gas in the swept-up shell.

An average jet power $\approx 7 \times 10^{38}$ 
erg s$^{-1}$ is almost an order of magnitude 
higher than the jet power in the most X-ray luminous 
Galactic microquasar, GRS 1915$+$105, during its 
long-term VHS oscillation cycles. The corresponding 
X-ray source, NGC\,5408 X-1, is itself 
almost an order of magnitude more luminous than GRS 1915$+$105.
On the other hand, an average jet power $\sim 10^{39}$ 
erg s$^{-1}$ has been inferred for SS\,433 and for 
the ULX NGC\,1313 X-2, both of which exhibit 
an extended lobe structure.
We briefly summarized the accretion states in which a jet 
may be produced, and the relation between jet power, 
X-ray luminosity and Eddington luminosity. 
Our results are consistent with a luminosity 
$L_{\rm X} \approx L_{\rm Edd} \approx 10^{40}$ erg s$^{-1}$ 
(BH mass $\approx 100 M_{\odot}$), 
and an average jet power $L_{\rm J} \sim 0.1 L_{\rm X}$; 
they also suggest that the ULX has been 
at least moderately active 
(accretion power $\ga 10^{-2} L_{\rm Edd}$) for most 
of the past $10^5$ yr.

Understanding this radio source in NGC\,5408  
is crucial for constraining the nature of ULXs, 
their association with starburst environments, 
their process of accretion, 
and in particular the balance between 
radiative luminosity emitted by the accretion disk 
and mechanical luminosity carried the jet. 
Very few ULXs have a detected radio counterpart.
Two ULXs in M\,82 may have radio counterparts: 
one is probably a SNR, while the nature 
of the other one is still unclear 
(K\"{o}rding et al.~2005);  
however, the physical association between 
those sources and the nearby ULXs
is still uncertain, given the density 
of sources in that region.
The only other unambiguous ULX-radio source 
association is in Holmberg II (Miller, 
Mushotzky \& Neff 2005). In that case, 
too, the spectrum is consistent with 
optically-thin synchrotron, though 
less steep than in our case ($\alpha_{\rm R} \approx -0.5$).
The Holmberg II radio source is resolved (size $\approx 50$ pc) 
and has a flux density $\approx 0.7$ mJy at 5 GHz 
(Miller et al.~2005), corresponding to $\approx 0.3$ mJy 
at the distance of NGC\,5408; hence, the two sources 
may be very similar (the X-ray luminosity and spectral 
properties of the corresponding ULXs 
are also similar: Dewangan et al.~2004; Soria et al.~2004). 
In that case, the pure SNR scenario appears 
inconsistent with the optical line spectrum; 
the most likely interpretation is that the synchrotron 
bubble is powered by the ULX jet (Miller et al.~2005).


Finally, should we expect to find  
many more microquasar lobes in nearby galaxies, 
perhaps misidentified as SNR? Taking NGC\,5408 and Holmberg II 
as typical cases, such lobes would require 
a jet injection power $\ga 10^{39}$ erg s$^{-1}$ 
over $\ga 10^{5}$ yr, in order to be detected by existing 
radio surveys of nearby galaxies.
If we assume that the jet power $L_{\rm J} \sim L_{\rm X} \la 
L_{\rm Edd}$, such systems should be associated 
with potential ULXs. In some cases, the X-ray source 
may not be currently active (or active at ULX level), 
yet there would still be a bright synchrotron radio 
source, formed by the integrated source activity over its lifetime. 
Therefore, finding the number ratio between microquasar lobes 
with or without a currently active ULX core could 
in principle provide a constraint on the average 
duty cycle of ULX microquasars. 
As an example, two bright radio sources, 
consistent with synchrotron bubbles 
with minimum-energies $> 10^{52}$ erg (hence, 
candidate microquasar lobes or hypernovae) have 
recently been discovered in NGC\,7424 (Soria et al.~2006).
One is associated with a transient ULX, the other has 
no corresponding X-ray source, down to a completeness 
limit $\approx 3 \times 10^{37}$ erg s$^{-1}$. 
Thus, one could speculate that the ULX responsible for 
latter radio source is currently in the off state.
Further discussion of these issues is beyond the scope 
of this paper.


\section*{Acknowledgments}
We thank Richard Hunstead and Ilana Klamer 
for their technical support at the ATCA telescope, 
above and beyond the call of duty. We are grateful 
to Manfred Pakull for drawing our attention to 
the existence of an H$\alpha$-emitting nebula at the ULX position. 
We also thank Geoff Bicknell, Chris Copperwheat, Mark Cropper, 
Zdenka Kunci\'{c}, Jasmina Lazendi\'{c}, 
Tom Pannuti and Doug Swartz 
for useful discussions and suggestions, 
and the anonymous referee for a number of improvements 
to our paper. RS acknowledges support from an OIF 
Marie Curie Fellowship.

\appendix

\section[]{Notes on the minimum-energy condition}

Two alternative formulations are used in the literature 
to estimate the minimum energy associated with a radio source 
(e.g., Tudose et al.~2006).
One choice (e.g., Pohl 1993, Bicknell 2005), 
arguably the more physical, is to assume an energy range 
$(\gamma_{\rm min},\gamma_{\rm max})$ for the relativistic 
electrons. In this framework, following Bicknell (2005), a useful 
way of expressing the minimum-energy magnetic field is:
\begin{eqnarray}
B_{\rm min} &=&  \frac{m_e}{e} \left[ \frac{p+1}{2} \, (1+k) \,
	C_1^{-1}(p) \, \frac{c}{m_e} \right]^{2/(p+5)} 
	\nonumber\\
	&\times&
	\left[h(p,\gamma_{\rm min}, \gamma_{\rm max})\,  
	\frac{I_{\nu} \nu^{-\alpha_{\rm R}}}{2r_{\rm c}}\right]^{2/(p+5)}
	\nonumber\\
	&\approx& 5.687 \times 10^8 \times (3.291 \times 10^{37})^{2/(p+5)} 
	\nonumber\\
	&\times& \left[ \frac{p+1}{2} \, (1+k) \,
	C_1^{-1}(p)\,h(p,\gamma_{\rm min}, \gamma_{\rm max}) \right]^{2/(p+5)} 
	\nonumber\\
	&\times& \left[\left(\frac{I_{\nu}/2r_{\rm c}}{{\rm erg~s}^{-1}
	{\rm~cm}^{-3}{\rm~Hz}^{-1}{\rm~sr}^{-1}}\right) \,
	\nu^{-\alpha_{\rm R}}\right]
	^{2/(p+5)} \nonumber\\
	&&\ \ {\rm Gauss},
\end{eqnarray}
where the energy spectrum of the electrons is $N(E)dE \sim E^{-p}dE$, 
$\alpha_{\rm R} \equiv (1-p)/2$, $m_e$ and $e$ are the electron mass 
and charge, $c$ the speed of light, 
$I_{\nu}$ the specific surface brightness (flux 
per angular beam area), $(2r_{\rm c})$ is the characteristic source diameter, 
and $k$ is the ratio between the energy of protons 
and electrons. We have assumed a filling factor of 1, for simplicity.
The functions 
\begin{equation}
h(p,\gamma_{\rm min}, \gamma_{\rm max}) = \frac{1}{p-2}\,
	\left[\gamma_{\rm min}^{(2-p)} - \gamma_{\rm max}^{(2-p)}
	\right],
\end{equation}
\begin{eqnarray}
C_1(p) &=& \frac{3^{p/2}}{2^{(p+13)/2}\,\pi^{(p+2)/2}} \noindent\\
	&\times& \frac{\Gamma\left(\frac{p}{4}+\frac{19}{12}\right) \,
	\Gamma\left(\frac{p}{4}-\frac{1}{12}\right) \,
	\Gamma\left(\frac{p}{4}+\frac{1}{4}\right)}
	{\Gamma\left(\frac{p}{4}+\frac{7}{4}\right)},	
\end{eqnarray}
and $\Gamma(z)$ is the Gamma function.
The corresponding total (minimum) energy density is (Bicknell 2005):
\begin{eqnarray}
\epsilon_{\rm min} &=& \epsilon_e + \epsilon_B = 
	\left(\frac{4}{p+1} + 1\right)
	\left(\frac{B_{\rm min}^2}{2\mu_0}\right)\nonumber\\
	&\approx& 3.98 \times 10^{-3} \left(\frac{5+p}{p+1}\right)
	\left(\frac{B_{\rm min}}{{\rm Gauss}}\right)^2 \ {\rm erg~cm}^{-3}
\end{eqnarray}
In our analysis, we adopted 
$\gamma_{\rm min} \sim 1$, $\gamma_{\rm max} \sim 10^4$.
For a steep spectrum, the minimum energy depends only 
very weakly on the high-energy cut-off. As for the 
low-energy cut-off, there are theoretical arguments and observational 
evidence for the presence of low-energy electrons 
in a radio lobe (e.g., Blundell \& Rawlings 2000, their 
Section 11.1).

The alternative choice 
(e.g., Burbidge 1956; Pacholczyk 1970; Longair 1994; 
Willott et al.~1999) is to assume a frequency 
range $(\nu_{\rm min},\nu_{\rm max})$ of the radio emission.
In that case, the magnetic field for which 
the energy density is minimized is (Longair 1994):
\begin{eqnarray}
B_{\rm min} &\approx& 119.7 \, C_2(p)^{4/7} (1 + k)^{2/7}  
	\left(\frac{V}{{\rm cm}^3}\right)^{-2/7}\nonumber\\
	&\times& 
	\left(\frac{L_{\nu}}
	{{\rm erg~s}^{-1}{\rm~Hz}^{-1}}\right)^{2/7} \ \ {\rm Gauss,}
\end{eqnarray}
where $V$ is the volume of the bubble, and we have assumed a unit 
filling factor. The functions
\begin{eqnarray}
C_1(p) &\approx& 2.344 \times 10^{25} \, 
	\frac{(7.413 \times 10^{-19})^{(2-p)}}
	{(1.253 \times 10^{37})^{(p-1)/2}}\nonumber\\
	&\times&
	\frac{\left[\nu_{\rm min}^{(2-p)/2} 
	- \nu_{\rm max}^{(2-p)/2}\right] \,\nu^{(p-1)/2}}{a(p) (p-2)},	
\end{eqnarray}
\begin{equation}
a(p) = \frac{\sqrt \pi}{2(p+1)}\,
	\frac{\Gamma\left(\frac{p}{4}+\frac{19}{12}\right) \,
	\Gamma\left(\frac{p}{4}-\frac{1}{12}\right) \,
	\Gamma\left(\frac{p}{4}+\frac{5}{4}\right)}
	{\Gamma\left(\frac{p}{4}+\frac{7}{4}\right)},
\end{equation}
and $\Gamma(z)$ is again the Gamma function.
The minimum energy density is then (e.g., Tudose et al.~2006):
\begin{eqnarray}
\epsilon_{\rm min} &\approx& 1.33 \times 10^9 \, C_2(p)^{4/7} (1 + k)^{4/7}  
	\left(\frac{V}{{\rm cm}^3}\right)^{-4/7}\nonumber\\
	&\times& 
	\left(\frac{L_{\nu}}
	{{\rm erg~s}^{-1}{\rm~Hz}^{-1}}\right)^{4/7} \ {\rm erg~cm}^{-3}.
\end{eqnarray}

The physical connection between the two formalisms is that 
for an electron of energy $\gamma$, the peak of its synchrotron 
emission is at a frequency
\begin{equation}
\nu \approx 0.29 \times \frac{3eB}{4\pi m_e}\,\gamma^2 
	\approx 1.22 \times 10^6 \left(\frac{B}{{\rm Gauss}}\right) 
	\, \gamma^2 \ {\rm Hz}.
\end{equation}
In some cases (e.g., Willott et al.~1999; Punsly 2005), 
the minimum-energy quantities are estimated for 
a conventional lower cut-off at $\nu_{\rm min} = 10$ MHz 
(lowest frequency accessible to Earth-based radio observations), 
and then the energy density is corrected by a numerical
factor $f > 1$ to account for low-energy electrons 
emitting below 10 MHz, and for hadronic contributions. 
In our case, setting a low-frequency 
limit of 10 MHz would imply an implausibly high 
$\gamma_{\rm min} \approx 300$ for a leptonic plasma.
Empirically, it was shown (Blundell \& Rawlings 2000; 
Punsly 2005) that the energy in the lobes of radio galaxies 
is $\approx 10$--$20$ times the minimum-energy estimated 
from a leptonic plasma with a 10-MHz cut-off. 
For a radio spectral index $\approx -1$, our choice of 
a leptonic plasma with 
$\gamma_{\rm min} \approx 1$, $\gamma_{\rm max} \approx 10^4$ 
corresponds to a correction factor $f \approx 15$, 
consistent with the observations (see also Section 4)

\section[]{Time evolution of the radio bubble}

We outline here the derivation of Equation (5) from 
Equations (3,4) and the minimum-energy condition.
Adopting Falle (1991)'s self-similar solution, 
the radius $r_{\rm c}$ of an expanding cocoon 
scales as
\begin{equation}
r_{\rm c} \propto \rho(r)^{-1/(5-\zeta)} {\mathcal L}^{1/(5-\zeta)} 
	t^{3/(5-\zeta)}
\end{equation}
where the density $\rho(r) \sim \rho_0 r^{-\zeta}$. 
For a constant ISM density ($\zeta = 0$), Eq.(B1) 
reduces to Eq.(3). The bubble size is also related 
to the total energy $E_{\rm tot} = V \epsilon_{\rm tot} 
\sim r_{\rm c}^3 \epsilon_{\rm tot}$, and 
from Eq.(4):
\begin{equation}
r_{\rm c} \propto {\mathcal L}^{1/3} t^{1/3} \epsilon_{\rm tot}^{1/3},
\end{equation}
In the minimum-energy approximation, 
$\epsilon_{\rm tot} = \epsilon_{\rm min}$.
Hence, from Equations (A1,A5), 
\begin{equation}
\epsilon_{\rm tot} \propto r_{\rm c}^{-12/(p+5)}
	\left[F_{\nu} d^2 \nu^{(p-1)/2}\right]^{4/(p+5)}
\end{equation}
With simple algebraic manipulations, the system of equations 
(B1,B2,B3) can be solved as:
\begin{equation}
F_{\nu} \propto \rho(r)^{\frac{3+3p}{4(5-\zeta)}} \,
	{\mathcal L}^{\frac{22-5\zeta +2p -p\zeta}{4(5-\zeta)}}\, 
	t^{\frac{16-5\zeta -4p -p\zeta}{4(5-\zeta)}} \,
	\nu^{\frac{1-p}{2}} \, d^{-2},
\end{equation}
which is identical to Eq.(5) for $\zeta = 0$ 
and $p = 3$. Similarly to Heinz (2002) (who 
discussed the case $p=2$), the normalization coefficient 
in front of Eq.(B4) was fixed to be consistent with  
Willott et al.~(1999)'s study of radio lobe 
properties in radio galaxies (Section 4).

\end{document}